\PassOptionsToPackage{dvipsnames,svgnames,x12names}{xcolor}
\documentclass[10pt,a4paper,onecolumn]{article}
\usepackage{marginnote}
\usepackage{graphicx}
\usepackage{xcolor}
\usepackage{etoolbox}
\usepackage[affil-it]{authblk}
\usepackage{orcidlink}
\usepackage{titlesec}
\usepackage{calc}
\usepackage{tikz}
\usepackage{hyperref}
\usepackage{caption}
\usepackage{tcolorbox}
\usepackage{amssymb,amsmath}
\usepackage{ifxetex,ifluatex}
\usepackage{seqsplit}
\usepackage{xstring}

\usepackage{float}
\let\origfigure\figure
\let\endorigfigure\endfigure
\renewenvironment{figure}[1][2] {
    \expandafter\origfigure\expandafter[H]
} {
    \endorigfigure
}

\NewDocumentCommand\citeproctext{}{}
\NewDocumentCommand\citeproc{mm}{%
  \begingroup\def\citeproctext{#2}\cite{#1}\endgroup}
\makeatletter
 \let\@cite@ofmt\@firstofone
 \def\@biblabel#1{}
 \def\@cite#1#2{{#1\if@tempswa , #2\fi}}
\makeatother
\newlength{\cslhangindent}
\newlength{\csllabelwidth}
\newenvironment{CSLReferences}[2] 
 {\begin{list}{}{%
  \setlength{\itemindent}{0pt}
  \setlength{\leftmargin}{0pt}
  \setlength{\parsep}{0pt}
  \ifodd #1
   \setlength{\leftmargin}{\cslhangindent}
   \setlength{\itemindent}{-1\cslhangindent}
  \fi
  \setlength{\itemsep}{#2\baselineskip}}}
 {\end{list}}
\usepackage{calc}

\ifLuaTeX
\usepackage[bidi=basic]{babel}
\else
\usepackage[bidi=default]{babel}
\fi
\babelprovide[main,import]{american}

\DeclareUnicodeCharacter{2265}{\ensuremath{\geq}}   
\DeclareUnicodeCharacter{0308}{\"{}}

\let\textttOrig=\texttt
\def\texttt#1{\expandafter\textttOrig{\seqsplit{#1}}}
\renewcommand{\seqinsert}{\ifmmode
  \allowbreak
  \else\penalty6000\hspace{0pt plus 0.02em}\fi}

\makeatletter
\let\href@Orig=\href
\def\href@Urllike#1#2{\href@Orig{#1}{\begingroup
    \def\Url@String{#2}\Url@FormatString
    \endgroup}}
\def\href@Notdoi#1#2{\def\tempa{#1}\def\tempb{#2}%
  \ifx\tempa\tempb\relax\href@Urllike{#1}{#2}\else
  \href@Orig{#1}{#2}\fi}
\def\href#1#2{%
  \IfBeginWith{#1}{https://doi.org}%
  {\href@Urllike{#1}{#2}}{\href@Notdoi{#1}{#2}}}
\makeatother

\usepackage[top=3.5cm, bottom=3cm, right=1.5cm, left=1.0cm,
            headheight=2.2cm, reversemp, includemp, marginparwidth=4.5cm]{geometry}

\titleformat{\section}
  {\normalfont\sffamily\Large\bfseries}
  {}{0pt}{}
\titleformat{\subsection}
  {\normalfont\sffamily\large\bfseries}
  {}{0pt}{}
\titleformat{\subsubsection}
  {\normalfont\sffamily\bfseries}
  {}{0pt}{}
\titleformat*{\paragraph}
  {\sffamily\normalsize}

\usepackage{fancyhdr}
\makeatletter
\let\ps@plain\ps@fancy
\fancyheadoffset[L]{4.5cm}
\fancyfootoffset[L]{4.5cm}

\definecolor{linky}{rgb}{0.0, 0.5, 1.0}

\newtcolorbox{repobox}
   {colback=red, colframe=red!75!black,
     boxrule=0.5pt, arc=2pt, left=6pt, right=6pt, top=3pt, bottom=3pt}

\newcommand{\ExternalLink}{%
   \tikz[x=1.2ex, y=1.2ex, baseline=-0.05ex]{%
       \begin{scope}[x=1ex, y=1ex]
           \clip (-0.1,-0.1)
               --++ (-0, 1.2)
               --++ (0.6, 0)
               --++ (0, -0.6)
               --++ (0.6, 0)
               --++ (0, -1);
           \path[draw,
               line width = 0.5,
               rounded corners=0.5]
               (0,0) rectangle (1,1);
       \end{scope}
       \path[draw, line width = 0.5] (0.5, 0.5)
           -- (1, 1);
       \path[draw, line width = 0.5] (0.6, 1)
           -- (1, 1) -- (1, 0.6);
       }
   }

\patchcmd{\@maketitle}{center}{flushleft}{}{}
\patchcmd{\@maketitle}{center}{flushleft}{}{}
\patchcmd{\@maketitle}{\LARGE}{\LARGE\sffamily}{}{}
\def\maketitle{{%
  
  \AB@maketitle}}
\makeatletter
\renewcommand\AB@affilsepx{ \protect\Affilfont}
\renewcommand\AB@affilnote[1]{{\bfseries #1}\hspace{3pt}}
\renewcommand{\affil}[2][]%
   {\newaffiltrue\let\AB@blk@and\AB@pand
      \if\relax#1\relax\def\AB@note{\AB@thenote}\else\def\AB@note{#1}%
        \setcounter{Maxaffil}{0}\fi
        \begingroup
        \let\href=\href@Orig
        \let\texttt=\textttOrig
        \let\protect\@unexpandable@protect
        \def\thanks{\protect\thanks}\def\footnote{\protect\footnote}%
        \@temptokena=\expandafter{\AB@authors}%
        {\def\\{\protect\\\protect\Affilfont}\xdef\AB@temp{#2}}%
         \xdef\AB@authors{\the\@temptokena\AB@las\AB@au@str
         \protect\\[\affilsep]\protect\Affilfont\AB@temp}%
         \gdef\AB@las{}\gdef\AB@au@str{}%
        {\def\\{, \ignorespaces}\xdef\AB@temp{#2}}%
        \@temptokena=\expandafter{\AB@affillist}%
        \xdef\AB@affillist{\the\@temptokena \AB@affilsep
          \AB@affilnote{\AB@note}\protect\Affilfont\AB@temp}%
      \endgroup
       \let\AB@affilsep\AB@affilsepx
}
\makeatother

\renewcommand\Affilfont{\sffamily\small\mdseries}
\usepackage[T1]{fontenc}
\usepackage[utf8]{inputenc}

\IfFileExists{upquote.sty}{\usepackage{upquote}}{}
\IfFileExists{microtype.sty}{%
\usepackage{microtype}
\UseMicrotypeSet[protrusion]{basicmath} 
}{}

\usepackage{hyperref}
\hypersetup{unicode=true,
            pdftitle={Synthesizer: Synthetic Observables For Modern Astronomy},
            pdfborder={0 0 0},
            breaklinks=true,
            pdfauthor={Will J. Roper, Christopher Lovell, Aswin Vijayan, Stephen Wilkins, Hollis Akins, Sabrina Berger, Connor Sant Fournier, Thomas Harvey, Kartheik Iyer, Marco Leonardi, Sophie Newman, Borja Pautasso, Ashley Perry, Louise Seeyave, and Laura Sommovigo},
            pdflang={en-UK},
            colorlinks=true,
            linkcolor={Maroon},
            filecolor={Maroon},
            citecolor=linky,
            urlcolor=linky,
            colorlinks,breaklinks=true,
            urlcolor=[rgb]{0.0, 0.5, 1.0},
            linkcolor=[rgb]{0.0, 0.5, 1.0}}
\let\addcontentslineOrig=\addcontentsline
\def\addcontentsline#1#2#3{\bgroup
  \let\texttt=\textttOrig\addcontentslineOrig{#1}{#2}{#3}\egroup}
\let\markbothOrig\markboth
\def\markboth#1#2{\bgroup
  \let\texttt=\textttOrig\markbothOrig{#1}{#2}\egroup}
\let\markrightOrig\markright
\def\markright#1{\bgroup
  \let\texttt=\textttOrig\markrightOrig{#1}\egroup}

\IfFileExists{parskip.sty}{%
\usepackage{parskip}
}{
\setlength{\parindent}{0pt}
\setlength{\parskip}{6pt plus 2pt minus 1pt}
}
\providecommand{\tightlist}{%
  \setlength{\itemsep}{0pt}\setlength{\parskip}{0pt}}
\ifx\paragraph\undefined\else
\let\oldparagraph\paragraph
\renewcommand{\paragraph}[1]{\oldparagraph{#1}\mbox{}}
\fi
\ifx\subparagraph\undefined\else
\let\oldsubparagraph\subparagraph
\renewcommand{\subparagraph}[1]{\oldsubparagraph{#1}\mbox{}}
\fi

\title{Synthesizer: Synthetic Observables For Modern Astronomy}

\author[1%
  *%
  \ensuremath\mathparagraph]{Will J. Roper%
    \,\orcidlink{0000-0002-3257-8806}\,%
    }
\author[2%
  *%
  ]{Christopher Lovell%
    \,\orcidlink{0000-0001-7964-5933}\,%
    }
\author[1%
  *%
  ]{Aswin Vijayan%
    \,\orcidlink{0000-0002-1905-4194}\,%
    }
\author[1%
  *%
  ]{Stephen Wilkins%
    \,\orcidlink{0000-0003-3903-6935}\,%
    }
\author[3%
  ]{Hollis Akins%
    \,\orcidlink{0000-0003-3596-8794}\,%
    }
\author[4%
  ]{Sabrina Berger%
    \,\orcidlink{0000-0002-4064-7883}\,%
    }
\author[5%
  ]{Connor Sant Fournier%
    \,\orcidlink{0009-0004-0771-4476}\,%
    }
\author[6%
  ]{Thomas Harvey%
    \,\orcidlink{0000-0002-4130-636X}\,%
    }
\author[7%
  ]{Kartheik Iyer%
    \,\orcidlink{0000-0001-9298-3523}\,%
    }
\author[8%
  ]{Marco Leonardi%
    \,\orcidlink{0009-0008-3592-7830}\,%
    }
\author[2%
  ]{Sophie Newman%
    \,\orcidlink{0009-0001-3422-3048}\,%
    }
\author[1%
  ]{Borja Pautasso%
    \,\orcidlink{0009-0003-1839-591X}\,%
    }
\author[1%
  ]{Ashley Perry%
    \,\orcidlink{0009-0001-0739-6162}\,%
    }
\author[1%
  ]{Louise Seeyave%
    \,\orcidlink{0000-0002-7020-3079}\,%
    }
\author[9%
  ]{Laura Sommovigo%
    \,\orcidlink{0000-0002-2906-2200}\,%
    }

\affil[1]{Astronomy Centre, University of Sussex, Falmer, Brighton BN1
9QH, UK%
  }
\affil[2]{Institute of Cosmology and Gravitation, University of
Portsmouth, Burnaby Road, Portsmouth, PO1 3FX, UK%
  }
\affil[3]{Department of Astronomy, The University of Texas at Austin,
Austin, TX 78712, USA%
  }
\affil[4]{School of Physics, University of Melbourne, Parkville, VIC
3010, Australia%
  }
\affil[5]{Institute of Space Sciences and Astronomy, University of
Malta, Msida MSD 2080, Malta%
  }
\affil[6]{Jodrell Bank Centre for Astrophysics, University of
Manchester, Oxford Road, Manchester M13 9PL, UK%
  }
\affil[7]{Columbia Astrophysics Laboratory, Columbia University, 550
West 120th Street, New York, NY 10027, USA%
  }
\affil[8]{Leiden Observatory, Leiden University, PO Box 9513, NL-2300 RA
Leiden, The Netherlands%
  }
\affil[9]{Center for Computational Astrophysics, Flatiron Institute, 162
5th Ave, New York, NY 10010, USA\newline%
  }

\affil[$\mathparagraph$]{Corresponding author.%
}
\affil[*]{These authors contributed equally.%
}

\begin{document}
\maketitle

\marginpar{

  \begin{flushleft}
  \sffamily\small

  \vspace{2mm}

  {\bfseries Software}
  \begin{itemize}
    \setlength\itemsep{0em}
    \item \href{https://github.com/openjournals/joss-reviews/issues/8418}{\color{linky}{Review}} \ExternalLink
    \item \href{https://github.com/synthesizer-project/synthesizer}{\color{linky}{Repository}} \ExternalLink
  \end{itemize}

  \vspace{2mm}

  \par\noindent\hrulefill\par

  \vspace{2mm}

  {\bfseries Editor:} TBC \\
  \vspace{1mm}
    {\bfseries Reviewers:}
  \begin{itemize}
  \setlength\itemsep{0em}
    \item TBC
    \end{itemize}
    \vspace{2mm}

  {\bfseries Submitted:} 17 June 2025\\
  {\bfseries Published:} N/A

  \vspace{2mm}
  {\bfseries License}\\
  Authors of papers retain copyright and release the work under a Creative Commons Attribution 4.0 International License (\href{http://creativecommons.org/licenses/by/4.0/}{\color{linky}{CC BY 4.0}}).

  \end{flushleft}
}

\vspace{-0.8cm}
\section{Summary}\label{summary}

Synthesizer is a fast, flexible, modular, and extensible Python package
that empowers astronomers to turn theoretical galaxy models into
realistic synthetic observations - including spectra, photometry,
images, and spectral cubes - with a focus on interchangeable modelling
assumptions. By offloading computationally intensive tasks to threaded
C++ extensions, Synthesizer delivers both simplicity and speed, enabling
rapid forward-modelling workflows without requiring users to manage
low-level data processing and computational details.

\section{Statement of need}\label{statement-of-need}

Comparing theoretical models of galaxy formation with observations
traditionally relies on two main approaches, both translating
theoretical models into the observer space (a technique known as forward
modelling). The first uses computationally expensive dust radiative
transfer codes (e.g. \citeproc{ref-SKIRT}{Camps \& Baes, 2015};
\citeproc{ref-SUNRISE}{Jonsson, 2006};
\citeproc{ref-POWDERDAY}{Narayanan et al., 2021}); these codes are
typically computationally expensive, prioritising fidelity. The second
uses simpler, bespoke pipelines that sacrifice some physical fidelity to
generate observables rapidly from large datasets (e.g.
\citeproc{ref-Fortuni2023}{Fortuni et al., 2023};
\citeproc{ref-Marshall2025}{Marshall et al., 2025};
\citeproc{ref-FLARESIV}{Roper et al., 2022};
\citeproc{ref-FLARESII}{Vijayan et al., 2020};
\citeproc{ref-SYNTHOBS}{Wilkins et al., 2020}).

Simplified inverse modelling approaches, such as SED fitting (e.g.
\citeproc{ref-EAZY}{Brammer et al., 2008};
\citeproc{ref-BAGPIPES}{Carnall et al., 2018};
\citeproc{ref-PROSPECTOR}{Johnson et al., 2021}) work in the opposite
direction, translating observables into intrinsic physical quantities.
However, these methods can introduce biases and uncertainties from both
observational effects and model assumptions. Compounding these
uncertainties is the fact that converged inverse modelling techniques
are costly in their own right, necessitating a simplified parameter
space to ensure convergence in a reasonable time. Forward modelling is
therefore becoming increasingly important not only for probing the
validity of theoretical models, but also for quantifying the
uncertainties in the modelling assumptions themselves.

However, existing forward modelling tools often lack the flexibility to
explore modelling uncertainties, the usability and modularity to explore
a wide range of modelling assumptions, and the performance necessary to
explore a large parameter space and process modern-day large datasets.
Furthermore, they frequently lack comprehensive documentation, hindering
consistency, and reproducibility across a range of datasets.

Synthesizer addresses these shortcomings by offering:

\begin{itemize}
\item
  Flexibility: Anything that could be changed by the user is explicitly
  designed to be variable (for a quantitative model parameter) or
  exchangeable (for a qualitative modelling choice). This means that
  users can easily vary everything in a reproducible way, without
  needing to modify the core code.
\item
  Performance: Computationally intensive operations are optimised by
  employing C extensions with OpenMP threading. Without this
  performance, the aforementioned flexibility is moot; only by coupling
  flexibility with the performance to utilise it can we explore large,
  high-dimensional parameter spaces in a reasonable time.
\item
  Modularity: Synthesizer is object-oriented, with a focus on decoupled
  classes that can be specialised and then swapped out at will. This
  modularity, in conjunction with a reliance on templating and
  dependency injection (see Emission Models below), is what enables
  Synthesizer's flexibility, as well as its application to a diverse
  range of astrophysical problems in both forward and inverse modelling
\item
  Extensibility: Extensive documentation and a clear API enable users to
  extend the package with their own calculations, parameterisations and
  subclasses. From the beginning, Synthesizer has been designed to be
  expanded to fit the needs of all users, even as astronomy and
  astrophysics evolve.
\end{itemize}

Synthesizer's design facilitates apples-to-apples comparisons between
simulations and observations (e.g. \citeproc{ref-FLARESXVIII}{Wilkins et
al., 2025}), permits exhaustive tests of the impact of parameter choices
(e.g. \citeproc{ref-LTU-ILI}{Ho et al., 2024}), enables the forward
modelling of large datasets previously considered impractical (e.g.
\citeproc{ref-LTU-LOVELL}{Lovell et al., 2024}), and promotes open and
reproducible science.

\section{Package overview}\label{package-overview}

Synthesizer is structured around a set of core abstractions. Here we
give a brief outline of these abstractions and a link to the
documentation for each.

\begin{itemize}
\tightlist
\item
  \textbf{Components}: Represent
  \href{https://synthesizer-project.github.io/synthesizer/galaxy_components/stars.html}{stars},
  \href{https://synthesizer-project.github.io/synthesizer/galaxy_components/gas.html}{gas},
  and
  \href{https://synthesizer-project.github.io/synthesizer/galaxy_components/blackholes.html}{black
  holes}, encapsulating physical properties, and emission and emission
  generation methods. For more details, see the
  \href{https://synthesizer-project.github.io/synthesizer/galaxy_components/galaxy_components.html\#components}{components
  documentation}.
\item
  \textbf{Galaxies}: Combine multiple components into a single object,
  allowing for cohesive calculations with all components, taking account
  of their interdependencies. For more details, see the
  \href{https://synthesizer-project.github.io/synthesizer/galaxy_components/galaxy_components.html\#the-galaxy-object}{galaxies
  documentation}.
\item
  \textbf{Emission Grids}: N-dimensional lookup tables of precomputed
  spectra and lines. Precomputed grids are available for stellar
  population synthesis models, including BC03
  (\citeproc{ref-bc03}{Bruzual \& Charlot, 2003}), BPASS
  (\citeproc{ref-bpass}{Stanway \& Eldridge, 2018}), FSPS (Conroy et al.
  (\citeproc{ref-fsps1}{2009}), Conroy \& Gunn
  (\citeproc{ref-fsps2}{2010})), Maraston (Maraston
  (\citeproc{ref-maraston05}{2005}), Newman et al.
  (\citeproc{ref-newman25}{2025})), all reprocessed using Cloudy
  (\citeproc{ref-cloudy}{Ferland et al., 1998}). Grids of AGN emission
  can also be calculated and explored. Users can generate custom grids
  via the accompanying
  \href{https://github.com/synthesizer-project/grid-generation}{grid-generation
  package}. For more details, see the
  \href{https://synthesizer-project.github.io/synthesizer/emission_grids/grids.html}{grids
  documentation}.
\item
  \textbf{Emission Models}: Modular templates defining the process of
  producing emissions from components. These models can be used to
  extract, generate, transform, or combine emissions. These are the
  backbone of Synthesizer's flexibility and modularity. For more
  details, see the
  \href{https://synthesizer-project.github.io/synthesizer/emission_models/emission_models.html}{emission
  models documentation}.
\item
  \textbf{Emissions}: The output of combining components with an
  emission model. These emissions are either spectra stored in
  \href{https://synthesizer-project.github.io/synthesizer/emissions/emission_objects/sed_example.html}{\texttt{Sed}
  objects}, or line emissions stored in
  \href{https://synthesizer-project.github.io/synthesizer/emissions/emission_objects/lines_example.html}{\texttt{LineCollection}
  objects}.
\item
  \textbf{Instruments}: Definitions of filters, resolutions, PSFs, and
  noise models to convert emissions into photometry, spectroscopy,
  images, and data cubes. For more details, see the
  \href{https://synthesizer-project.github.io/synthesizer/observatories/observatories.html}{instruments
  documentation} and
  \href{https://synthesizer-project.github.io/synthesizer/observatories/filters.html}{filters
  documentation}.
\item
  \textbf{Observables}: Containers for the output spectra with
  observational effects
  (\href{https://synthesizer-project.github.io/synthesizer/observables/spectroscopy/spectroscopy.html}{\texttt{Sed}
  objects}), photometry
  (\href{https://synthesizer-project.github.io/synthesizer/observables/photometry/photometry.html}{\texttt{PhotometryCollection}
  objects}), images
  (\href{https://synthesizer-project.github.io/synthesizer/observables/imaging/imaging.html}{\texttt{Image}
  and \texttt{ImageCollection} objects}), and spectral data cubes
  (\href{https://synthesizer-project.github.io/synthesizer/observables/spectral_data_cubes/spectral_data_cubes.html}{\texttt{SpectralDataCube}
  objects}).
\end{itemize}

Synthesizer is hosted on
\href{https://github.com/synthesizer-project/synthesizer}{GitHub} and is
available on \href{https://pypi.org/project/cosmos-synthesizer/}{PyPI}.
The documentation is available through
\href{https://synthesizer-project.github.io/synthesizer/}{GitHub Pages}.

\subsection{Related packages}\label{related-packages}

Several packages either overlap with Synthesizer's functionality or
complement it in end-to-end workflows:

\begin{itemize}
\tightlist
\item
  \textbf{SPS \& photoionisation}: Libraries for stellar
  spectra---\textbf{BC03} (\citeproc{ref-bc03}{Bruzual \& Charlot,
  2003}), \textbf{FSPS} (\citeproc{ref-fsps1}{Conroy et al., 2009};
  \citeproc{ref-fsps2}{Conroy \& Gunn, 2010}), \textbf{BPASS}
  (\citeproc{ref-bpass}{Stanway \& Eldridge, 2018}), \textbf{Maraston}
  (\citeproc{ref-maraston05}{Maraston, 2005})---paired with dust/nebular
  models, plus \textbf{Cloudy} (\citeproc{ref-cloudy}{Ferland et al.,
  1998}) or \textbf{MAPPINGS} (\citeproc{ref-MAPPINGS}{Dopita \&
  Sutherland, 1996}) for reprocessing.
\item
  \textbf{Monte Carlo RT}: Photon--dust/gas simulators like
  \textbf{SKIRT} (\citeproc{ref-SKIRT}{Camps \& Baes, 2015}),
  \textbf{Powderday} (\citeproc{ref-POWDERDAY}{Narayanan et al., 2021}),
  \textbf{Hyperion} (\citeproc{ref-hyperion}{Robitaille, 2011}). These
  robust forward-modelling codes produce SEDs and images ingestible by
  Synthesizer.
\item
  \textbf{PSF \& instrument tools}: \textbf{STPSF}
  (\citeproc{ref-stpsf}{Perrin et al., 2014}) (JWST, Roman, HST) and
  \textbf{GalSim} (\citeproc{ref-galsim}{Rowe et al., 2015}) model
  telescope optics, detector effects, and noise.
\item
  \textbf{Pre/post-processing}: \textbf{YT} (\citeproc{ref-YT}{Turk et
  al., 2011}) for volumetric data analysis and visualization of
  simulation outputs; \textbf{Astroquery}
  (\citeproc{ref-astroquery}{Ginsburg et al., 2019}) for automated
  querying of astronomical archives and catalogs; \textbf{Dense Basis}
  (\citeproc{ref-dense_basis}{Iyer et al., 2019}) offers
  SED-/SFH-tailored basis functions.
\item
  \textbf{Inverse modeling \& SED fitting}: \textbf{EAZY}
  (\citeproc{ref-EAZY}{Brammer et al., 2008}), \textbf{BAGPIPES}
  (\citeproc{ref-BAGPIPES}{Carnall et al., 2018}), \textbf{PROSPECTOR}
  (\citeproc{ref-PROSPECTOR}{Johnson et al., 2021}) extract galaxy
  properties from SEDs.
\end{itemize}

\section{Acknowledgements}\label{acknowledgements}

We acknowledge the use of the following software packages in this work:
\href{https://www.astropy.org/}{Astropy} (\citeproc{ref-astropy}{Astropy
Collaboration et al., 2022}),
\href{https://unyt.readthedocs.io/en/stable/index.html}{unyt}
(\citeproc{ref-unyt}{Goldbaum et al., 2018}),
\href{https://matplotlib.org/}{Matplotlib}
(\citeproc{ref-matplotlib}{Hunter, 2007}),
\href{https://numpy.org/}{NumPy} (\citeproc{ref-numpy}{Harris et al.,
2020}), \href{https://www.scipy.org/}{SciPy}
(\citeproc{ref-scipy}{Virtanen et al., 2020}), and
\href{https://www.openmp.org/}{OpenMP} (\citeproc{ref-openmp}{Dagum \&
Menon, 1998}).

WJR, APV, and SMW acknowledge support from the Sussex Astronomy Centre
STFC Consolidated Grant (ST/X001040/1). CCL acknowledges support from a
Dennis Sciama fellowship funded by the University of Portsmouth for the
Institute of Cosmology and Gravitation. APV acknowledges support from
the Carlsberg Foundation (grant no CF20-0534). SB is supported by the
Melbourne Research Scholarship and N D Goldsworthy Scholarship. LS and
SN are supported by an STFC studentship. This work was supported by the
Simons Collaboration on ``Learning the Universe''.

This work used the DiRAC@Durham facility managed by the Institute for
Computational Cosmology on behalf of the STFC DiRAC HPC Facility
(www.dirac.ac.uk). The equipment was funded by BEIS capital funding via
STFC capital grants ST/K00042X/1, ST/P002293/1, ST/R002371/1 and
ST/S002502/1, Durham University and STFC operations grant ST/R000832/1.
DiRAC is part of the National e-Infrastructure.

\section*{References}\label{references}
\addcontentsline{toc}{section}{References}

\phantomsection\label{refs}


\begin{CSLReferences}{1}{0}
\bibitem[\citeproctext]{ref-astropy}
Astropy Collaboration, Price-Whelan, A. M., Lim, P. L., Earl, N.,
Starkman, N., Bradley, L., Shupe, D. L., Patil, A. A., Corrales, L.,
Brasseur, C. E., Nöthe, M., Donath, A., Tollerud, E., Morris, B. M.,
Ginsburg, A., Vaher, E., Weaver, B. A., Tocknell, J., Jamieson, W.,
\ldots{} Astropy Project Contributors. (2022). {The Astropy Project:
Sustaining and Growing a Community-oriented Open-source Project and the
Latest Major Release (v5.0) of the Core Package}. \emph{935}(2), 167.
\url{https://doi.org/10.3847/1538-4357/ac7c74}

\bibitem[\citeproctext]{ref-EAZY}
Brammer, G. B., van Dokkum, P. G., \& Coppi, P. (2008). {EAZY: A Fast,
Public Photometric Redshift Code}. \emph{686}(2), 1503--1513.
\url{https://doi.org/10.1086/591786}

\bibitem[\citeproctext]{ref-bc03}
Bruzual, G., \& Charlot, S. (2003). {Stellar population synthesis at the
resolution of 2003}. \emph{344}(4), 1000--1028.
\url{https://doi.org/10.1046/j.1365-8711.2003.06897.x}

\bibitem[\citeproctext]{ref-SKIRT}
Camps, P., \& Baes, M. (2015). {SKIRT: An advanced dust radiative
transfer code with a user-friendly architecture}. \emph{Astronomy and
Computing}, \emph{9}, 20--33.
\url{https://doi.org/10.1016/j.ascom.2014.10.004}

\bibitem[\citeproctext]{ref-BAGPIPES}
Carnall, A. C., McLure, R. J., Dunlop, J. S., \& Davé, R. (2018).
{Inferring the star formation histories of massive quiescent galaxies
with BAGPIPES: evidence for multiple quenching mechanisms}.
\emph{480}(4), 4379--4401. \url{https://doi.org/10.1093/mnras/sty2169}

\bibitem[\citeproctext]{ref-fsps2}
Conroy, C., \& Gunn, J. E. (2010). {The Propagation of Uncertainties in
Stellar Population Synthesis Modeling. III. Model Calibration,
Comparison, and Evaluation}. \emph{712}(2), 833--857.
\url{https://doi.org/10.1088/0004-637X/712/2/833}

\bibitem[\citeproctext]{ref-fsps1}
Conroy, C., Gunn, J. E., \& White, M. (2009). {The Propagation of
Uncertainties in Stellar Population Synthesis Modeling. I. The Relevance
of Uncertain Aspects of Stellar Evolution and the Initial Mass Function
to the Derived Physical Properties of Galaxies}. \emph{699}(1),
486--506. \url{https://doi.org/10.1088/0004-637X/699/1/486}

\bibitem[\citeproctext]{ref-openmp}
Dagum, L., \& Menon, R. (1998). OpenMP: An industry standard API for
shared-memory programming. \emph{Computational Science \& Engineering,
IEEE}, \emph{5}(1), 46--55. \url{https://doi.org/10.1109/99.660313}

\bibitem[\citeproctext]{ref-MAPPINGS}
Dopita, M. A., \& Sutherland, R. S. (1996). {Spectral Signatures of Fast
Shocks. I. Low-Density Model Grid}. \emph{102}, 161.
\url{https://doi.org/10.1086/192255}

\bibitem[\citeproctext]{ref-cloudy}
Ferland, G. J., Korista, K. T., Verner, D. A., Ferguson, J. W., Kingdon,
J. B., \& Verner, E. M. (1998). {CLOUDY 90: Numerical Simulation of
Plasmas and Their Spectra}. \emph{110}(749), 761--778.
\url{https://doi.org/10.1086/316190}

\bibitem[\citeproctext]{ref-Fortuni2023}
Fortuni, F., Merlin, E., Fontana, A., Giocoli, C., Romelli, E.,
Graziani, L., Santini, P., Castellano, M., Charlot, S., \& Chevallard,
J. (2023). {FORECAST: A flexible software to forward model cosmological
hydrodynamical simulations mimicking real observations}. \emph{677},
A102. \url{https://doi.org/10.1051/0004-6361/202346725}

\bibitem[\citeproctext]{ref-astroquery}
Ginsburg, A., Sipőcz, B. M., Brasseur, C. E., Cowperthwaite, P. S.,
Craig, M. W., Deil, C., Guillochon, J., Guzman, G., Liedtke, S., Lian
Lim, P., Lockhart, K. E., Mommert, M., Morris, B. M., Norman, H.,
Parikh, M., Persson, M. V., Robitaille, T. P., Segovia, J.-C., Singer,
L. P., \ldots{} a subset of the astropy collaboration. (2019).
{astroquery: An Astronomical Web-querying Package in Python}.
\emph{157}, 98. \url{https://doi.org/10.3847/1538-3881/aafc33}

\bibitem[\citeproctext]{ref-unyt}
Goldbaum, N. J., ZuHone, J. A., Turk, M. J., Kowalik, K., \& Rosen, A.
L. (2018). Unyt: Handle, manipulate, and convert data with units in
python. \emph{Journal of Open Source Software}, \emph{3}(28), 809.
\url{https://doi.org/10.21105/joss.00809}

\bibitem[\citeproctext]{ref-numpy}
Harris, C. R., Millman, K. J., Walt, S. J. van der, Gommers, R.,
Virtanen, P., Cournapeau, D., Wieser, E., Taylor, J., Berg, S., Smith,
N. J., Kern, R., Picus, M., Hoyer, S., Kerkwijk, M. H. van, Brett, M.,
Haldane, A., Río, J. F. del, Wiebe, M., Peterson, P., \ldots{} Oliphant,
T. E. (2020). Array programming with {NumPy}. \emph{Nature},
\emph{585}(7825), 357--362.
\url{https://doi.org/10.1038/s41586-020-2649-2}

\newpage

\bibitem[\citeproctext]{ref-LTU-ILI}
Ho, M., Bartlett, D. J., Chartier, N., Cuesta-Lazaro, C., Ding, S.,
Lapel, A., Lemos, P., Lovell, C. C., Makinen, T. L., Modi, C., Pandya,
V., Pandey, S., Perez, L. A., Wandelt, B., \& Bryan, G. L. (2024).
{LtU-ILI: An All-in-One Framework for Implicit Inference in Astrophysics
and Cosmology}. \emph{The Open Journal of Astrophysics}, \emph{7}, 54.
\url{https://doi.org/10.33232/001c.120559}

\bibitem[\citeproctext]{ref-matplotlib}
Hunter, J. D. (2007). Matplotlib: A 2D graphics environment.
\emph{Computing in Science \& Engineering}, \emph{9}(3), 90--95.
\url{https://doi.org/10.1109/MCSE.2007.55}

\bibitem[\citeproctext]{ref-dense_basis}
Iyer, K. G., Gawiser, E., Faber, S. M., Ferguson, H. C., Kartaltepe, J.,
Koekemoer, A. M., Pacifici, C., \& Somerville, R. S. (2019).
Nonparametric star formation history reconstruction with gaussian
processes. I. Counting major episodes of star formation. \emph{The
Astrophysical Journal}, \emph{879}(2), 116.
\url{https://doi.org/10.3847/1538-4357/ab2052}

\bibitem[\citeproctext]{ref-PROSPECTOR}
Johnson, B. D., Leja, J., Conroy, C., \& Speagle, J. S. (2021). {Stellar
Population Inference with Prospector}. \emph{254}(2), 22.
\url{https://doi.org/10.3847/1538-4365/abef67}

\bibitem[\citeproctext]{ref-SUNRISE}
Jonsson, P. (2006). {SUNRISE: polychromatic dust radiative transfer in
arbitrary geometries}. \emph{372}(1), 2--20.
\url{https://doi.org/10.1111/j.1365-2966.2006.10884.x}

\bibitem[\citeproctext]{ref-LTU-LOVELL}
Lovell, C. C., Starkenburg, T., Ho, M., Anglés-Alcázar, D., Davé, R.,
Gabrielpillai, A., Iyer, K., Matthews, A. E., Roper, W. J., Somerville,
R., Sommovigo, L., \& Villaescusa-Navarro, F. (2024). {Learning the
Universe: Cosmological and Astrophysical Parameter Inference with Galaxy
Luminosity Functions and Colours}. \emph{arXiv e-Prints},
arXiv:2411.13960. \url{https://doi.org/10.48550/arXiv.2411.13960}

\bibitem[\citeproctext]{ref-maraston05}
Maraston, C. (2005). {Evolutionary population synthesis: models,
analysis of the ingredients and application to high-z galaxies}.
\emph{362}(3), 799--825.
\url{https://doi.org/10.1111/j.1365-2966.2005.09270.x}

\bibitem[\citeproctext]{ref-Marshall2025}
Marshall, M. A., Amen, L., Woods, T. E., Côté, P., Yung, L. Y. A.,
Amenouche, M., Pass, E. K., Balogh, M. L., Salim, S., \& Moutard, T.
(2025). {FORECASTOR - II. Simulating galaxy surveys with the
Cosmological Advanced Survey Telescope for Optical and UV Research}.
\emph{537}(2), 1703--1719. \url{https://doi.org/10.1093/mnras/staf065}

\bibitem[\citeproctext]{ref-POWDERDAY}
Narayanan, D., Turk, M. J., Robitaille, T., Kelly, A. J., McClellan, B.
C., Sharma, R. S., Garg, P., Abruzzo, M., Choi, E., Conroy, C., Johnson,
B. D., Kimock, B., Li, Q., Lovell, C. C., Lower, S., Privon, G. C.,
Roberts, J., Sethuram, S., Snyder, G. F., \ldots{} Wise, J. H. (2021).
{POWDERDAY: Dust Radiative Transfer for Galaxy Simulations}.
\emph{252}(1), 12. \url{https://doi.org/10.3847/1538-4365/abc487}

\bibitem[\citeproctext]{ref-newman25}
Newman, S. L., Lovell, C. C., Maraston, C., Giavalisco, M., Roper, W.
J., Saxena, A., Vijayan, A. P., \& Wilkins, S. M. (2025).
{Cloudy-Maraston: Integrating nebular continuum and line emission with
the Maraston stellar population synthesis models}. \emph{arXiv
e-Prints}, arXiv:2501.03133.
\url{https://doi.org/10.48550/arXiv.2501.03133}

\bibitem[\citeproctext]{ref-stpsf}
Perrin, M. D., Sivaramakrishnan, A., Lajoie, C.-P., Elliott, E., Pueyo,
L., Ravindranath, S., \& Albert, Loı̈c. (2014). {Updated point spread
function simulations for JWST with WebbPSF}. In J. M. Oschmann Jr., M.
Clampin, G. G. Fazio, \& H. A. MacEwen (Eds.), \emph{Space telescopes
and instrumentation 2014: Optical, infrared, and millimeter wave} (Vol.
9143, p. 91433X). \url{https://doi.org/10.1117/12.2056689}

\bibitem[\citeproctext]{ref-hyperion}
Robitaille, T. P. (2011). {HYPERION: an open-source parallelized
three-dimensional dust continuum radiative transfer code}. \emph{536},
A79. \url{https://doi.org/10.1051/0004-6361/201117150}

\bibitem[\citeproctext]{ref-FLARESIV}
Roper, W. J., Lovell, C. C., Vijayan, A. P., Marshall, M. A., Irodotou,
D., Kuusisto, J. K., Thomas, P. A., \& Wilkins, S. M. (2022). First
light and reionisation epoch simulations (flares) -- IV. The size
evolution of galaxies at z~≥~5. \emph{Monthly Notices of the Royal
Astronomical Society}, \emph{514}(2), 1921--1939.
\url{https://doi.org/10.1093/mnras/stac1368}

\bibitem[\citeproctext]{ref-galsim}
Rowe, B. T. P., Jarvis, M., Mandelbaum, R., Bernstein, G. M., Bosch, J.,
Simet, M., Meyers, J. E., Kacprzak, T., Nakajima, R., Zuntz, J.,
Miyatake, H., Dietrich, J. P., Armstrong, R., Melchior, P., \& Gill, M.
S. S. (2015). {GALSIM: The modular galaxy image simulation toolkit}.
\emph{Astronomy and Computing}, \emph{10}, 121--150. 
\url{https://doi.org/10.1016/j.ascom.2015.02.002}

\bibitem[\citeproctext]{ref-bpass}
Stanway, E. R., \& Eldridge, J. J. (2018). {Re-evaluating old stellar
populations}. \emph{479}(1), 75--93.
\url{https://doi.org/10.1093/mnras/sty1353}

\newpage

\bibitem[\citeproctext]{ref-YT}
Turk, M. J., Smith, B. D., Oishi, J. S., Skory, S., Skillman, S. W.,
Abel, T., \& Norman, M. L. (2011). {yt: A Multi-code Analysis Toolkit
for Astrophysical Simulation Data}. \emph{192}(1), 9.
\url{https://doi.org/10.1088/0067-0049/192/1/9}

\bibitem[\citeproctext]{ref-FLARESII}
Vijayan, A. P., Lovell, C. C., Wilkins, S. M., Thomas, P. A., Barnes, D.
J., Irodotou, D., Kuusisto, J., \& Roper, W. J. (2020). {First Light And
Reionization Epoch Simulations (FLARES) -- II: The photometric
properties of high-redshift galaxies}. \emph{Monthly Notices of the
Royal Astronomical Society}, \emph{501}(3), 3289--3308.
\url{https://doi.org/10.1093/mnras/staa3715}

\bibitem[\citeproctext]{ref-scipy}
Virtanen, P., Gommers, R., Oliphant, T. E., Haberland, M., Reddy, T.,
Cournapeau, D., Burovski, E., Peterson, P., Weckesser, W., Bright, J.,
van der Walt, S. J., Brett, M., Wilson, J., Millman, K. J., Mayorov, N.,
Nelson, A. R. J., Jones, E., Kern, R., Larson, E., \ldots{} SciPy 1.0
Contributors. (2020). {{SciPy} 1.0: Fundamental Algorithms for
Scientific Computing in Python}. \emph{Nature Methods}, \emph{17},
261--272. \url{https://doi.org/10.1038/s41592-019-0686-2}

\bibitem[\citeproctext]{ref-SYNTHOBS}
Wilkins, S. M., Lovell, C. C., Fairhurst, C., Feng, Y., Matteo, T. D.,
Croft, R., Kuusisto, J., Vijayan, A. P., \& Thomas, P. (2020).
{Nebular-line emission during the Epoch of Reionization}. \emph{Monthly
Notices of the Royal Astronomical Society}, \emph{493}(4), 6079--6094.
\url{https://doi.org/10.1093/mnras/staa649}

\bibitem[\citeproctext]{ref-FLARESXVIII}
Wilkins, S. M., Vijayan, A. P., Hagen, S., Caruana, J., Conselice, C.
J., Done, C., Hirschmann, M., Irodotou, D., Lovell, C. C., Matthee, J.,
Plat, A., Roper, W. J., \& Taylor, A. J. (2025). {First Light and
Reionization Epoch Simulations (FLARES) -- XVIII: the ionising
emissivities and hydrogen recombination line properties of early AGN}.
\emph{arXiv e-Prints}, arXiv:2505.05257.
\url{https://doi.org/10.48550/arXiv.2505.05257}

\end{CSLReferences}
\end{document}